\documentclass[a4paper,11pt]{amsart}
\begin{document}
\hyphenation{li-nea-rized peri-astron phy-si-cal-ly ener-gy
ina-de-quate}
\title[] {{\bf  No GW is emitted by B PSR1913+16}}
\author[]{Angelo Loinger}
\thanks{To be
published on \emph{Spacetime \& Substance.} \\email:
angelo.loinger@mi.infn.it\\ Dipartimento di Fisica, Universit\`a
di Milano, Via Celoria, 16 - 20133 Milano (Italy)}

\begin{abstract}
In the \emph{exact} (non-linear) formulation of general relativity
(GR) \emph{no} motion of bodies can give origin to gravitational
waves (GW's) -- as it has been proved. Accordingly, the measured
rate of change of the orbital period of binary pulsar B PSR1913+16
must have \emph{other} causes, different from the emission of
GW's; maybe the viscous losses of the unseen pulsar companion, if
it were e.g. a helium star.
\end{abstract}

\maketitle

\vskip1.20cm
\textbf{Summary}. - The paper is structured as follows.
Sect.\textbf{1}. contains some passages of a recent report by
Weisberg and Taylor (see $[1g)]$ regarding thirty years of
observations of binary radiopulsar B PSR1913+16. Sect.\textbf{2}.:
a straightforward criticism of the relativistic approach which is
employed in papers cited in \cite{1}. Sect.\textbf{2bis}.: a
possible alternative explanation of the shrinkage of the orbit of
the above radiopulsar. Sect.\textbf{3}.: the linear approximation
of GR is \emph{inadequate} to give an existence theorem of
physical GW's. Sect.\textbf{4}.: \emph{system B PSR1913+16 cannot
emit GW's}. Sect.\textbf{5}.: the analogy between Maxwell-Lorentz
e.m. theory and the linearized version of GR is a \emph{false}
analogy. Sect.\textbf{6}.: erroneousness of a surmise concerning
the behaviour of B PSR1913+16.

\vskip1.20cm
\textbf{1}. -  Nobody has ever found a \emph{\textbf{direct}}
experimental proof of the \emph{real}  existence of the
gravitational waves (GW's). According to some authors \cite{1}, an
\emph{\textbf{indirect}} experimental evidence could be given by
the time decrease of the orbital period $P_{b}$ of the binary
pulsar B PSR1913+16 \cite{1bis}.

\par The abstract of paper $[1g)]$ runs as follows: ``We
describe results derived from thirty years of observations of PSR
B1913+16. Together with the Keplerian orbital parameters,
measurements of the relativistic periastron advance and a
combination of gravitational redshift and time dilation yield the
stellar masses with high accuracy. The measured rate of change of
orbital period agrees with that expected from the emission of
gravitational radiation, according to general relativity, to
within about $0.2$ percent. Systematic effects depending on the
pulsar distance and on poorly known galactic constants now
dominate the error budget, so tighter bounds will be difficult to
obtain. $[\ldots]$.''. And in sect.\textbf{3}.\textbf{1} of the
same paper $[1g)]$ the authors claim that: ``According to general
relativity, a binary star system should emit energy in the form of
gravitational waves. The loss of orbital energy results in
shrinkage of the orbit, which is most easily observed as a
decrease in orbital period. Peters and Mathews (1963) (see
\cite{2}) showed that in general relativity the rate of period
decrease is given by

\begin{eqnarray} \label{eq:one}
\dot{P}_{b,GR} & = & - \frac{192 \pi G^{5/3}}{5c^{5}}
\left(\frac{P_{b}}{2 \pi} \right)^{-5/3}
\left(1-e^{2}\right)^{-7/2} \times {} \nonumber\\
& & {}
\left(1+\frac{73e^{2}}{24}+\frac{37e^{4}}{96}\right)m_{p}m_{c}
\left(m_{p}+m_{c}\right)^{-1/3}.''
\end{eqnarray}

Here: $G$ is the gravitational constant; $c$ the speed of light
\emph{in vacuo}; $e$ the orbital eccentricity $(e=0.6171338(4))$;
$m_{p}$ the mass of the pulsar ($m_{p}=1.4414 \pm0.0002$ solar
masses), $m_{c}$ the mass of the companion ($m_{c}=1.3867
\pm0.0002$ solar masses).

\par Then, Weisberg and Taylor $[1g)]$ write:
``Comparison of the measured $\dot{P}_{b}$ with the theoretical
value requires a small correction, $\dot{P}_{b,Gal}$, for relative
acceleration between the solar system and binary pulsar system,
projected onto the line of sight (Damour and Taylor 1991) $[see
[1e)]]$. This correction is applied to the measured $\dot{P}_{b}$
to form a ``corrected value'' $\dot{P}_{b,
corrected}=\dot{P}_{b}-\dot{P}_{b,Gal}$. The correction term
depends on several rather poorly known quantities, including the
distance $[\approx16,000$ light-years$]$ and proper motion of the
pulsar and the radius of the Sun's galactic orbit. The best
currently available values yield
$\dot{P}_{b,Gal}=-(0.0128\pm0.0050)\times 10^{-12}$ s/s, so that
$\dot{P}_{b, corrected}=2.4056\pm0.0051)\times 10^{-12}$ s/s.
Hence

\begin{equation} \label{eq:two}
    \frac{\dot{P}_{b, corrected}}{\dot{P}_{b,GR}} = 1.0013 \pm
    0.0021 ,
\end{equation}

and we conclude that the measured orbital decay is consistent at
the $(0.13 \pm 0.21)$ \% level with the general relativistic
prediction for the emission of gravitational radiation.
$[\ldots]$.''

\vskip0.50cm
\textbf{2}. -  The good agreement between the measured
$\dot{P}_{b}$ and the computed $\dot{P}_{b}$ is suspect -- as I
have already emphasized \cite{3} --, because the relativistic
perturbative approximation, of which eq.(\ref{eq:one}) is a
consequence, is quite unreliable from the point of view of the
\emph{\textbf{exact}} (non-linear) formulation of GR, as it was
pointed out by several relativists \cite{3bis}.

\par Further, I remark that in GR the hypothetic GW's do not have
a \emph{true} energy. Therefore the \emph{true} mechanical energy
which is lost during the orbital motion should transform itself
into the \emph{pseudo} (i.e. false) energy of the hypothetic GW's:
the energy balance would be violated. \emph{\textbf{Objection}}:
if we \emph{suppose} that the \emph{linearized} version of GR has
an unconditioned, approximate validity -- as the experimentalists,
and some (simple) theoreticians \cite{4}, do -- the physical
existence of GW's seems a theoretical possibility, and it seems --
by exploiting the analogy with Maxwell-Lorentz e.m. theory -- that
the \emph{acceleration} of a body can generate GW's.
\emph{\textbf{Answer}}: the energy-momentum of such GW's has a
tensor character only under Lorentz transformations of
co-ordinates, but \emph{not} under general transformations. Now,
this is contrary to the basic tenet of GR.

\vskip0.50cm
\textbf{2bis}. -  The authors of papers \cite{1} have
\emph{assumed} that \emph{both} stars of the considered binary
system are neutron stars, and thus act dynamically as \emph{point}
masses. But if the companion star were a helium star or a white
dwarf, tides and viscous actions might mimic the relativistic (as
the periastron advance) and pseudorelativistic effects. In
particular, the viscous losses of the companion could give a time
decrease of the pulsar revolution period of the same order of
magnitude of that given by the hypothesized emission of
gravitational radiation -- as it is well known to many
observational astrophysicists.

\par Finally, the empirical success of a theory -- or of a given
computation -- is not an absolute guaranty for its conceptual
adequacy. Consider for instance the Ptolemaic theory of cycles and
epicycles, which explained rather well the planetary orbits (with
the only exception of Mercury's).

\vskip0.50cm
\textbf{3}. -  It can be proved that the linear approximation of
GR is quite \emph{inadequate} to a proper study of the hypothetic
GW's, see \cite{5}, \cite{6}. And if we continue the approximation
beyond the linear stage (see \cite{7}, \cite{8}), we find that the
radiation terms of the gravitational field can be \emph{destroyed}
by convenient co-ordinate transformations: this proves that the
GW's are \emph{only a product of a special choice of the reference
system}, i.e. that they do not possess a \emph{physical} reality
(see further \cite{9}, \cite{10}, \cite{11}): the undulatory
solutions of Einstein field equations have a mere \emph{formal}
(\emph{non}-physical) character.

\vskip0.50cm
\textbf{4}. -  If the two stars of B PSR1913+16 are dynamically
treated as two (gravitationally interacting) \emph{point} masses
\cite{1}, the \emph{\textbf{exact}} formulation of GR tells us
that their orbits are \emph{\textbf{geodetic}} lines \cite{9},
i.e. their motions are ``natural'', ``free'' motions, quite
analogous to a rectilinear and uniform motion of a point charge in
the customary Maxwell-Lorentz theory. Accordingly, no GW is sent
forth by our stars! \cite{10}, \cite{11}.

\par In my paper $[3 \beta)]$ I have given another elegant proof of
this fact, resting on a fundamental proposition by Hermann Weyl
\cite{12}, according to which for any relative motion of two
bodies it is always possible (in GR) to choose a co-ordinate
system in which \emph{both} bodies are \emph{at rest}. (Remark
that in GR the expression \emph{at rest} must be defined precisely
by specifying the \emph{interested spacetime manifold}.)

\par Let us apply the above proposition to system B PSR1913+16,
i.e. let us choose a co-ordinate frame for which both stars are at
rest. Evidently, an observer $\Omega$ who ``resides'' in this
frame does not record any emission of GW's. Now, any observer
$\Omega'$ -- very far, in particular, from $\Omega$ --, for whom B
PSR1913+16 is in motion, does not possess (in GR!) any physical
privilege with respect to $\Omega$. Accordingly, both observers,
$\Omega$ and $\Omega'$, do not register any GW sent forth by our
binary system. (See Weyl \cite{13} for the Riemann-Einstein
manifold of two point masses at rest.)

\vskip0.50cm
\textbf{5}. -  The \emph{\textbf{false}} formal analogy between
the e.m. Maxwell-Lorentz theory and the linearized version of GR
is the responsible for the publication of countless and senseless
papers. In particular, it has generated the conviction that, in
GR, the \emph{acceleration} of a body must give origin to GW's;
many people have forgotten that, in the \emph{exact} (non-linear)
formulation of GR, the concept ``acceleration'' does not possess
an absolute character. (The above conviction was also extended to
perturbative approximations of higher order.)

\par We observe finally that the exact theory does not admit any
class of physically privileged reference frames for which, in
particular, the undulatory character of a given gravitational
field is an invariant property.

\vskip0.50cm
\textbf{6}. -  A last remark. Some authors have conjectured that a
coexistence of effects due to emission of GW's by B PSR1913+16 and
to tides and viscous actions of the companion star could be
possible.

\par Now, this is pure nonsense, because -- as it can be proved --
even motions that are \emph{not} purely gravitational cannot
generate GW's \cite{14}.

\small \vskip0.5cm\par\hfill {$\Pi$$\tilde{\upsilon}$$\rho$
$\sigma$$o$$\iota$
$\pi$$\rho$$o$$\sigma$$o$$\acute{\iota}$$\sigma$$\omega$.}
\vskip0.10cm\par\hfill {\emph{(I will bring fire to thee.)}}
     \vskip0.10cm\par\hfill {EURIPIDES, \emph{Andromache}.}

\vskip2.0cm
\begin{center}
$^{\star---------------------------\star}$
\end{center}


\begin{thebibliography}{999}
\bibitem[1bis]{1bis}
In his Nobel lecture -- see \emph{Revs. Mod. Phys.}, \textbf{66}
(1994) 699 -- R.A. Hulse does not discuss the question of the
GW's, which would be emitted by B PSR1913+16.
\end{thebibliography}
\end{document}